\documentclass[11pt,twoside]{article}
\usepackage{asp2010}

\resetcounters

\begin{document}

\title{Nucleosynthesis origin of PG 1159 stars, Sakurai's object
and of rare subclasses of presolar grains}
\author{R. Gallino$^{1,2}$, O. Straniero$^2$, E. Zinner$^3$, M.
Jadhav$^3$,   
L. Piersanti$^2$, S. Cristallo$^{2,4}$, \& S. Bisterzo$^1$
\affil{$^1$Dipartimento di Fisica Generale Universit{\`a} di Torino, Italy}
\affil{$^2$INAF, Osservatorio di Teramo, Italy}
\affil{$^3$Laboratory for Space Sciences and Department of Physics,
Washington University, St. Louis}
\affil{$^4$Departamento de Fisica Teorica y del Cosmos, Universidad de
Granada, Spain}}

\begin{abstract}
We discuss theoretical AGB predictions for 
hydrogen-deficient
PG 1159 stars and Sakurai's object, which show peculiar enhancements in He, C and O, and how these 
enhancements may be understood in the framework
of a very late thermal pulse nucleosynthetic event. We then discuss the nucleosynthesis origin of 
rare subclasses of presolar grains extracted from carbonaceous
meteorites, the SiC AB grains showing low  $^{12}$C/$^{13}$C in the range  2 to 10
and the very few high-density graphite grains with $^{12}$C/$^{13}$C around 10.
\end{abstract}

\section{H-deficient stars}
\subsection{PG 1159 stars: extremely hot Post-AGB stars}

The hydrogen-deficiency in extremely hot post-AGB stars of spectral class
PG 1159, which includes about 40 stars, with  $T_{\rm eff}$ ranging from
75\,000 and 200\,000 K and log $g$ from 5.5 to 7.5, is probably caused by 
a very late thermal pulse in the He shell (VLTP, Sch{\"o}nberner 1979, Iben 1984) 
while the post-AGB star is in the hot WD cooling sequence. 
Because of the high $T$ $_{\rm eff}$ in PG 1159 stars, all species are highly
ionized and, hence, most metals are only accessible by UV spectroscopy.
A passionate work has been conducted in the last 20 years by Klaus Werner and 
collaborators (Werner \& Herwig 2006, Werner et al. 2009, 2010 and references therein).
In Table 1 we report in particular the range of peculiar abundances of He, C, N, O 
estimated in PG 1159 stars, where the mass fraction of He ranges between
0.30  and 0.85, C between 0.15 and 0.40, N between 0.001 and 0.01, and O between 0.02 and 0.2. 
The energy released by the VLTP forces the stellar radius to inflate and
the star to cool and  proceed back toward a born-again AGB. At the maximum extension
of the convective thermal instability the very small residual and inactivated H shell
is likely engulfed by the pulse and severly depleted, 
so that the usually hidden He-, C-rich and s-rich He intershell
is eventually exposed to the photosphere. 
PG 1159 stars are seemingly descendants of [WC] stars, 
which show similar HeCNO peculiarities.
  
\subsection{Sakurai's object (V4334 Sgr)}
\begin{figure}
\includegraphics[angle=-90,width=6.5cm]{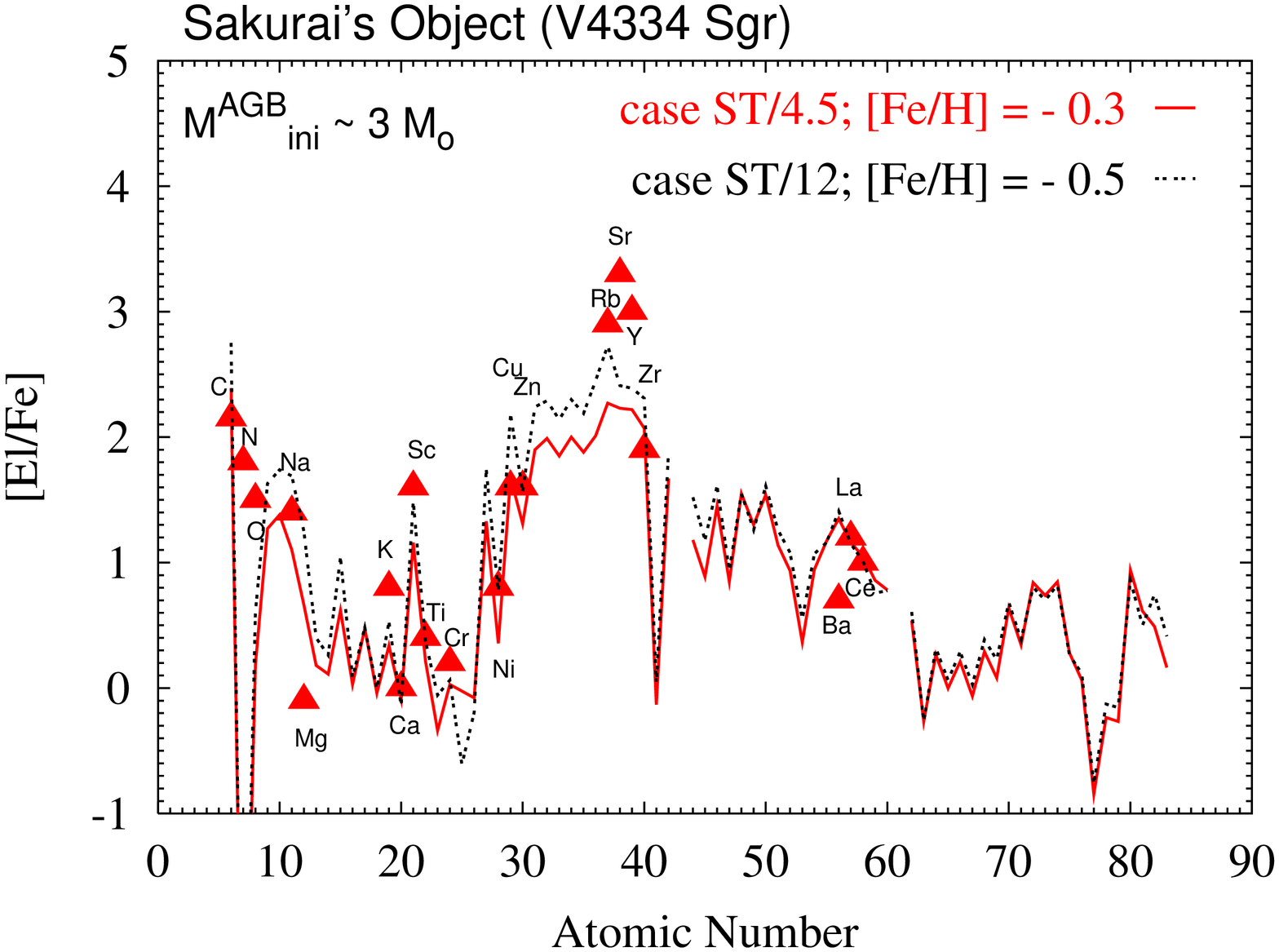}
\includegraphics[angle=-90,width=6.5cm]{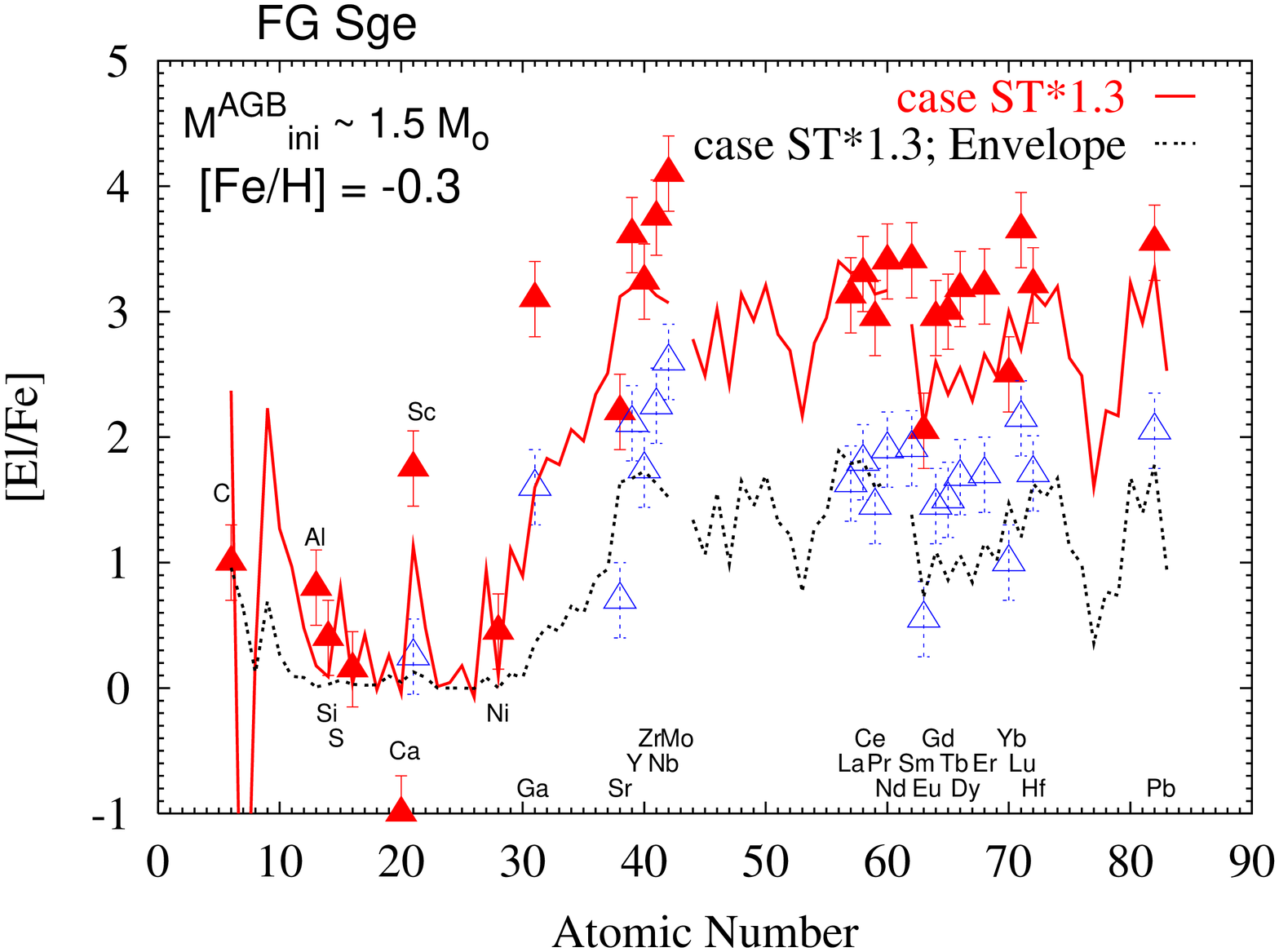}
\caption{Left panel:  Sakurai's object spectroscopic data by Asplund et al.
(1998)
compared with the [El/Fe] distribution in the He intershell for different AGB
models
of $M^{\rm AGB}_{ini}$ = 3 $M_\odot$, [Fe/H] = $-$0.3, case ST/4.5 at the last
thermal
pulse, or for [Fe/H] = $-$0.5 and case ST/12.
Right panel: FG Sge spectroscopic data by Gonzalez et al. (1998)
compared with  an AGB model of
$M^{\rm AGB}_{ini}$ = 1.5 $M_\odot$, [Fe/H] = $-$0.3, case ST $\times$ 1.3
at the last
thermal pulse.
The lower curve is for the distribution in the envelope at the last thermal
pulse for the same AGB model (see text).
}
\label{fig1}
\end{figure}

Sakurai's object (V4334 Sgr) was discovered in 1996 and soon recognized to
be the central star of an old planetary nebula (6000 yr ago). It recently 
underwent a VLTP and is now a hydrogen-deficient born-again AGB star.
The estimated mass fractions He/C/O = 0.90/0.07/0.03 and other element
abundances (Asplund et al. 1998, their Figure 6)
are reported in Table 1 and compared
with PG1159 stars. This composition is based on the choice of the best
adopted atmospheric model with  C/He $\approx$ 0.10 (by number). 
Note that there is a "carbon problem" for Sakurai's
object similar to
that encountered for the hydrogen-deficient RCrB stars. 
Indeed, the spectroscopic C abundance derived from CI lines is
about 0.6 dex smaller than the one deduced from the selected 
model atmosphere. The same is the case for [Fe/H].
However, the relative abundance ratios, like [El/Fe], are scarcely dependent 
on the C/He choice.
In Fig. 1 left panel we compare the observed [El/Fe] data of Sakurai's object
with predicted abundances in the He intershell at the last thermal pulse 
for an AGB model of $M^{\rm AGB}_{ini}$ = 3 $M_\odot$, [Fe/H] = $-$0.3, case ST/4.5. 
Given the uncertainty of the initial metallicity, we also plot in the figure a similar predicted
[El/Fe] distribution for $M^{\rm AGB}_{ini}$ = 3
$M_\odot$, [Fe/H] = $-$0.5 and case ST/12 (dashed line). In this case, Sc and Rb appear 
better reproduced, but the reverse is true for C and Cu. As to Sr, its 
abundance in October 1996 was 
overestimated. The presence of  carbon dust buffers around Sakurai's object
may in general introduce a noticeable uncertainty in spectroscopic
abundances. 
V 605 Aql (Nova Aql 1919) is a second star having likely  suffered a VLTP about 90 yr
ago. Clayton et al. (2006) estimated He/C/O = 0.54/0.40/0.05.
Both V106 Aql and Sakurai's object showed peculiar
rapid declines and fading characteristic of episodic carbon dust emission,
as in the case of R CrB stars.
However, as discussed below, several R CrB stars likely originated
in a completely different way, as binary WD mergers.

\subsection{H ingestion and partial burning in the He intershell}

In order to produce a consistent amount of N and to achieve the low
$^{12}$C/$^{13}$C observed in Sakurai's object, ingestion and burning of
hydrogen
in a TP  is impossible when the H shell is still active. It  may work when a
very thin H
envelope is left after the star leaves the AGB (Herwig et al. 1999,
Miller Bertolami et al. 2006).
A quite low $^{12}$C/$^{13}$C $\le$ 10 results while a consistent amount of
$^{14}$N is
built up.
During the AGB phase, standard elemental mass fractions in the He intershell after
a thermal pulse are He/C/O = 0.75/0.20/0.005.
Higher C and O abundances may be obtained by  including an
efficient overshoot at the base of the convective thermal
pulse in the TP-AGB phase (Herwig et al. 1997).
Alternatively, proper account should be given to the peeling effect by mass 
loss both at the tip of the AGB and in the early phase of the post-AGB
track. There, a "superwind" of up to several 
10$^{-5}$ $M_\odot$/yr has been measured. Lawlor \& MacDonald (2006)
introduced these effects in their stellar evolution code in an wide
spectrum of initial masses and metallicities.
Another important factor is the thickness of the He buffer, which  decreases
with increasing the CO core mass, that is with the initial stellar mass.
The authors showed that chemical peculiarities observed in stars
having suffered the VLTP do not strictly require overshoot in the AGB phase.
One should also consider that the bottom of the VLTP is degenerate,
different from what occurs during the AGB phase, with the possibility to further
increase C and O.

\begin{table*}
\caption{PG 1159 stars, Sakurai's object, AGB  He-intershell predictions,
extra needs. The spectroscopic data for Sakurai's objects are from Asplund et
al. (1998), their Fig. 6. As to Ne, our AGB predictions refer to $^{22}$Ne,
the most abundant isotope in the He intershell. The range of $^{19}$F for AGB He-intershell
predictions increases with the number of thermal pulses, i.e. with the initial AGB
mass.}

\label{table1}
\begin{center}
\resizebox{13cm}{!}{\begin{tabular}{llllll}
\\
\hline
                     &PG 1159 & Sakurai's object
& He intershell & extra needs       \\
$^{12}$C/$^{13}$C           & not measurable & 2 to 5  &
NO ${13}$C   &   H-b
\\
H                &deficient            &deficient &0.&
\\
He               & from 0.30 to 0.85   &  0.90     &  0.75
& partial He-b
\\
C                & from 0.15 to 0.60   &0.07     &  0.2
&partial He-b
\\
N               & from 0.001 to 0.01        & 0.01 &
0. &          H-b
\\
O               & from 0.02 to 0.20 & 0.03       &  0.005
& partial He-b
\\
F               & 1 to 250$\times$solar  & not detectable       & 1 to 250 &  
\\
Ne          & 0.02             & 0.02& 0.02 &
\\
Na             &&25$\times$solar & 8$\times$solar &
\\
Si              & from solar to 0.5 $\times$ solar   & 7$\times$solar  &
solar &
\\
S           & from solar to 0.1$\times$solar         &2.5$\times$solar  &  solar &
\\
P               & from solar to 2 $\times$ solar   &          &  solar to
4$\times$solar &
\\
Ar          & solar      &                           &  solar &
\\
Fe              & solar to 0.1 $\times$ solar & 0.6$\times$solar  &  solar to
0.7$\times$solar &  \\
Ni              & not enhanced      &   8$\times$solar  &  not enhanced &
\\
Cu              && 55$\times$solar &25$\times$solar&
\\
Zn              && 55$\times$solar &20$\times$solar&
\\ 
s-elements  & impossible to detect  & highly enhanced    &
highly enhanced&
\\
\hline
\end{tabular}}
\end{center}
\end{table*}

\subsection{FG Sge }
The peculiar FG Sge has also been assumed to have recently suffered a VLTP.
Over the last 120 years FG Sge evolving from a hot post-AGB star to a present cool 
and born-again AGB.   
Figure 1 (panel right)  shows the FG Sge spectroscopic data by 
Gonzalez et al. (1998) (full triangles)
compared with theoretical predictions (upper curve) 
in the  He-intershell after the last thermal pulse for an AGB
initial mass of 1.5 $M_\odot$, metallicity [Fe/H] = $-$0.3 and 
the $^{13}$C pocket choice ST $\times$ 1.3. Note the huge [ls/Fe], [hs/Fe]
and [Pb/Fe], of the order of 3 dex each.
The  high value [Eu/Fe] = 2 dex observed is s-process Eu, in agreement with
the typical s-process expectation [La/Fe]$_s$ $\approx$ 1 dex.
Moreover, Gonzalez et al. estimated $^{12}$C/$^{13}$C $>$ 10, and provided 
first spectroscopic evidence of H-deficiency.
Jeffery \& Sch{\"o}nberner (2006) reanalyzed all
extant spectroscopic data and atmospheric parameters, raising doubts 
on the huge s-process abundances derived by Gonzalez et al. 
(1998). The lower  $T_{\rm eff}$ = 5500 K chosen for the  model
atmosphere brought Jeffery \& Sch{\"o}nberner to conclude that the s-process 
element abundances are
more that one order of magnitude less than inferred, most likely inherited
already during its previous  AGB phase. They concluded that FG Sge suffered a
late thermal pulse (LTP), not a VLTP, then evolving back to
a born-again AGB. In the lower curve of Figure 1, we
compare the envelope AGB model prediction at the last thermal pulse. Note
that the two indicators of the s-process distribution,  [hs/ls] and [Pb/hs], would remain
unaltered. The spectroscopic heavy elements have been reduced by 1.5 dex
(empty triangles),
what corresponds to a typical factor of 30 of dilution of He intershell
material mixed with the envelope.  
Note that predicted [C/Fe] would better compare with a LTP solution.

\subsection{The R CrB  stars}
\begin{figure}
\includegraphics[angle=0,width=6cm]{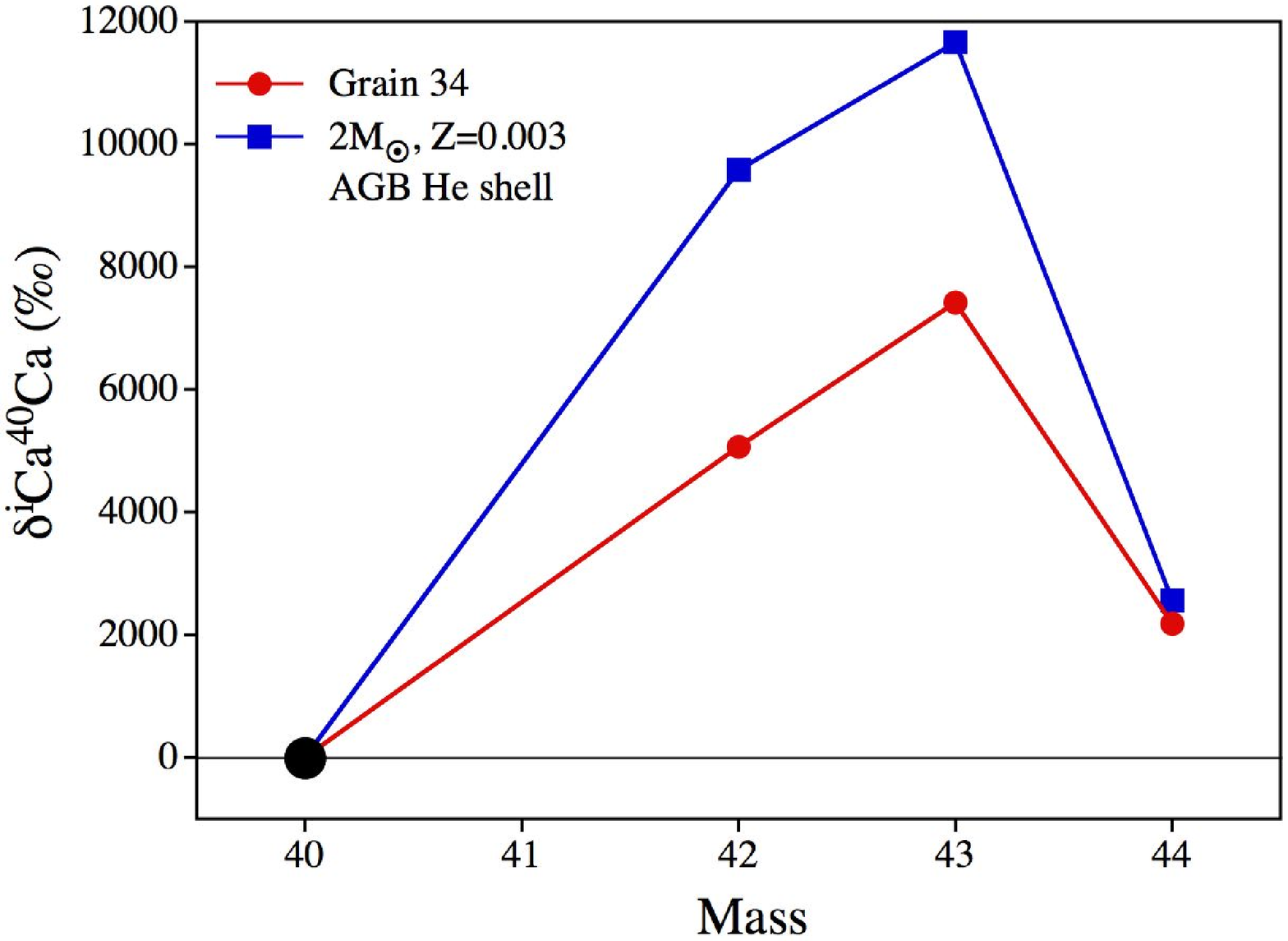}
\includegraphics[angle=0,width=6cm]{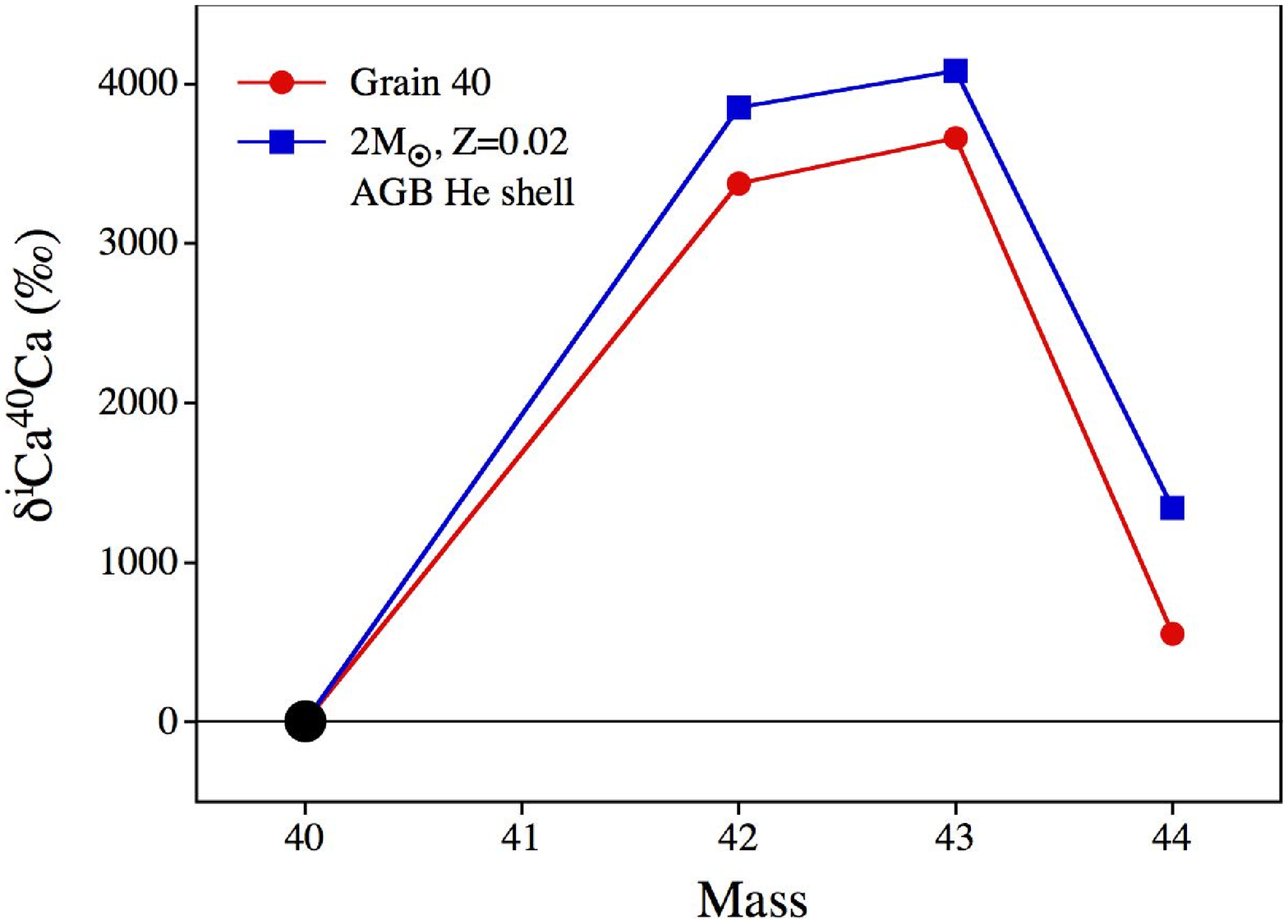}
\caption{Permil variation with respect to solar of Ca isotopes of graphite
presolar grains g-34 and g-40 compared with  He intershell predictions of
two different AGB models (adapted from Jadhav et al. 2008).
}
\label{fig2}  
\end{figure}  
So far about 50 R CrB have been discovered. Their atmospheres are extremely
hydrogen deficient and carbon rich. 
Another distinctive feature of some R CrB stars is the enormous F detection,
in the range 1,000 to 8,000 times solar for  [Fe/H]  in the
range  $-$0.5 to  $-$2.0 (Pandey et al. 2008). Such drastic
$^{18}$O and $^{19}$F excesses indicate that the  merging of a
CO-WD with a He-WD gives rise to partial He burning and production 
of $^{18}$O via  $\alpha$-capture on $^{14}$N, accompanied by
$^{18}$O(p,$\gamma$)$^{19}$F. Detailed nucleosynthesis calculations   
for these peculiar objects are not easy however.

\section{Presolar grains}

A subclass of presolar SiC grains discovered in carbonaceus
meteorites, the  SiC of type AB  (4 to 5 \% of all presolar SiC grains) are characterized by very low 
$^{12}$C/$^{13}$C ratios, in the range 2 to 10. Mainstream SiC (covering 93\% of all presolar SiC
grains), show higher $^{12}$C/$^{13}$C ratios, from $\sim$10 to 100 
(solar ratio is 89), averaging at around 60 (Zinner 1998, 2008). While
mainstream SiC likely originated in low mass AGB stars of around solar
metallicity, the stellar origin of SiC AB grains is still
enigmatic. These grains clearly show the signature of H burning
in the CNO cycle and H burning is also indicated
by their relatively high inferred $^{26}$Al/$^{27}$Al ratios (Amari
et al. 2001). However, the low $^{12}$C/$^{13}$C ratios are difficult
to reconcile with the condition C$>$O, necessary for
SiC condensation. J-type carbon stars and born-again AGBs
like  Sakurai's object have
been proposed as sources of AB grains. 
Despite  SiC AB grains show low $^{12}$C/$^{13}$C, the permil variations of $^{29}$Si/$^{28}$Si
and  $^{30}$Si/$^{28}$Si with respect to solar is indistinguishable from mainstream SiC
that reach maximum values of $\approx$200 and
$\approx$150, respectively. Instead, far higher permil
variations are predicted in the He intershell. 
This indicates that SiC AB grains are incompatible with an origin 
in born-again AGBs like Sakurai's object, unless one speculates that the grains
formed in a cool circumstellar disk, after having been mixed with
previously ejected material.

Very rare high-density graphite grains have been discovered with the signature
of He intershell in the trace elements Ca and Ti (Jadhav et al. 2008).
 A couple of examples are reported in the two panels of Fig. 3, for the
grains g-34 and g-40.  A similar exceptional permil variation has been detected
for both Ca and Ti in grain g-9. Also trace Mg and Si are present
but they show essentially normal isotopic composition, maybe related with isotopic 
equilibration with solar material.

\acknowledgements R.G. thanks the INAF-Osservatorio di Teramo for financial support.


\vspace{3mm}
\textsc{ \bf{References} }

\noindent Amari, S, et al. (2001), ApJ, 559, 643\\
Asplund, M., Gustafsson, B., Rao, N. K., \& Lambert, D. L. (1998), A\&A, 332,
651\\
Clayton, G. C., et al. (2006), ApJ, 646, L69\\ 
Gonzalez, G., et al. (1998), ApJS, 114, 133\\
Herwig, F., Bl{\"o}cker, T., Sch{\"o}nberner, D., \& El Eid, M. (1997), A\&A, 324,
L81\\ 	
Herwig, F., Blcker, T., Langer, N., \& Driebe, T. (1999), A\&A, 349, L5\\	
Iben, I., Jr (1984), ApJ, 277, 333\\
Jadhav, M., et al. (2008), ApJ, 682, 1479\\
Jeffery, C. S., \& Sch{\"o}nberner, D. (2006), A\&A, 459, 885\\
Miller Bertolami, M. M. et al. (2006),A\&A, 449, 313\\
Pandey, G., Lambert, D. L., \& Rao, N. K. (2008), ApJ, 674, 1068\\
Sch{\"o}nberner, D. (1979),  A\&A, 79, 108\\
Werner, K., \& Herwig, F., (2006), PASP, 118, 183\\
Werner, K., Rauch, T., \& Kruk, J. W. (2010), ApJ, 719, L32\\
Werner, K., et al. (2009), Astrophys. Space Sci., 320, 159 \\
Zinner, E. (1998), Annu. Rev. Earth Planet. Sci. 26, 147 \\
Zinner, E. (2008), PASA, 25, 7\\
%

\end{document}